\documentclass[twocolumn,showpacs,aip,prb,preprintnumbers,amsmath,amssymb,superscriptaddress,reprint]{revtex4}
\usepackage{graphicx}% Include figure files
\usepackage{dcolumn}% Align table columns on decimal point
\usepackage{bm}% bold math
\usepackage{color}
\usepackage{mathptmx}
\usepackage{epstopdf}

\DeclareMathOperator{\tr}{\mathop{\mathrm{Tr}}}

\newcommand{\mapright}[1]{\smash{\mathop{\hbox to 1.0cm{\rightarrowfill}}\limits^{#1}}}

\begin{document}

\preprint{}

\title{
\textcolor{blue}{Efficient electron refrigeration using superconductor/spin-filter devices}
}

\author{Shiro Kawabata}
\affiliation{Electronics and Photonics Research Institute (ESPRIT), National Institute of Advanced Industrial Science and Technology (AIST), Tsukuba, Ibaraki, 305-8568, Japan}

\author{Asier Ozaeta}
\affiliation{Centro de F\'{i}sica de Materiales (CFM-MPC), Centro Mixto CSIC-UPV/EHU, Manuel de Lardizabal 5, E-20018 San Sebasti\'{a}n, Spain}

\author{Andrey S. Vasenko}
\affiliation{LPMMC, Universit\'{e} Joseph Fourier and CNRS, 25 Avenue des Martyrs, BP 166, 38042 Grenoble, France}

\author{Frank W. J. Hekking}
\affiliation{LPMMC, Universit\'{e} Joseph Fourier and CNRS, 25 Avenue des Martyrs, BP 166, 38042 Grenoble, France, and Institut Universitaire de France, 103, bd Saint-Michel 75005 Paris, France}

\author{F. Sebasti\'{a}n Bergeret}
\affiliation{Centro de F\'{i}sica de Materiales (CFM-MPC), Centro Mixto CSIC-UPV/EHU, Manuel de Lardizabal 5, E-20018 San Sebasti\'{a}n, Spain}
\affiliation{Donostia International Physics Center (DIPC), Manuel de Lardizabal 4, E-20018 San Sebasti\'{a}n, Spain}

\date{\today}

\begin{abstract}
Efficient electron-refrigeration based on a
normal-metal/spin-filter/superconductor junction is proposed and demonstrated
theoretically. 
The spin-filtering effect leads to values of  the cooling power  
much higher than in conventional normal-metal/nonmagnetic-insulator/superconductor coolers and allows
 for an efficient extraction of heat from the normal metal.  
 We demonstrate that highly efficient cooling can be  realized in both ballistic and 
diffusive multi-channel junctions in which the reduction of the electron temperature from
300 mK to around 50 mK can be achieved. Our results indicate the practical usefulness of
 spin-filters for efficiently cooling detectors, sensors, and quantum
devices.
\end{abstract}

\maketitle

The flow of charge current in N/I/S (normal metal/insulator/ superconductor) tunnel junctions at a bias voltage $V$ is accompanied by a heat transfer from N into S.
This phenomenon arises due to presence of  the superconducting energy-gap $\Delta$ which allows for a
 selective tunneling of high-energy "hot" quasiparticles out of N.
Such a heat transfer through N/I/S junctions can be used for the realization of microcoolers.~\cite{rf:Nahum,rf:Leivon,rf:Giazotto06,rf:Muhonen}
Present state-of-the-art experiments allow the reduction of the electron temperature in a normal metal lead from 300 to about 100 mK, offering perspectives for on-chip cooling of nano or micro systems, such as high-sensitive sensors, detectors and quantum devices.~\cite{rf:Clark,rf:ONeil,rf:Lowell}

However,  a serious limitation of the performance of N/I/S microcoolers arises from the
intrinsic multi-particle nature of charge transport in N/I/S junctions which is governed
not only by single-particle tunneling but also by two-particle processes due to the
Andreev reflection.~\cite{rf:Giazotto06} While the single-particle current and the
associated heat current are due to quasiparticles with energies $E>\Delta$, at low
temperatures or high junction transparencies the charge transport in N/I/S junctions is
dominated by a subgap process: the Andreev reflection. 
In such a process, electrons with energies smaller than $\Delta$ are reflected as holes at the N/I/S interface,
leading to the transfer of a Cooper pair  into the superconductor. 
Since the energy of the electrons and holes involved in the process
are symmetric with respect to the Fermi energy, there is no heat current  through the interface.
However, by applying a subgap bias across the junctions the Andreev reflection results in a finite charge current $I_A$
flowing through the N/I/S system.
 Due to finite resistance of the normal metal, this current generates Joule heating $I_AV$, which is entirely deposited in the normal metal.~\cite{rf:Bardas,rf:Rajauria,rf:Vasenko} 
This heating exceeds the single-particle cooling  at temperatures low enough, and therefore the suppression of Andreev processes is desirable for an efficient cooling.

One way to decrease the Andreev current  is by  decreasing the N/I/S junction transparency.
However, large contact resistance hinders "hot" carrier transfer and { leads} to a
severe limitation in the achievable cooling powers. In order to increase the junction
transparency and at the same time to reduce the Andreev current, it was suggested to use
ferromagnetic metals (FM).~\cite{rf:Giazotto02,rf:Ozaeta} Giazotto and co-workers have
theoretically investigated the cooling of a clean  N/FM/S junction and predicted an
enhancement of the cooling power compared to N/I/S junctions due to the suppression of
the Andreev Joule heating.~\cite{rf:Giazotto02}  
In order to realize such an efficient cooler, however, challenging half-metallic FMs~\cite{rf:Wolf} with extremely-high spin-polarization $P > 0.94$  
are needed.

In the present work we theoretically propose an alternative N/I/S microcooler  
with  a ferromagnetic insulator as a tunneling barrier which acts  as a spin-filter. This was demonstrated 
  in experiments 
using europium chalcogenides tunneling barriers.~\cite{rf:Moodera1}
The spin-filtering effect suppresses   
the Andreev reflection in a N/spin-filter(SF)/S junction as the one shown in Fig. 1(a).~\cite{rf:Kashiwaya,rf:Bergeret2}
 We show that this suppression leads  in both, ballistic and  diffusive, N/SF/S junctions  to   dramatic enhancement of the cooling power  
 which gives rise
to a dramatic reduction of the final achievable electron temperature.

%
%
%
%\section{Ballistic junction}
%
%
%
%

In order to illustrate the basic cooling  mechanism using spin-filters we first consider
 a one-dimensional clean N/SF/S junction [Fig.~1(a)]. 
In the following we set $\hbar = 1$.
 The SF barrier can be modeled by a
spin-dependent delta-function potential [see  Fig.~1(a)], $i. e.$, $V_\sigma (x) = \left(
V + \rho_\sigma U \right) \delta(x)$, where $V$ is the spin-independent part of
the potential, $U$ is the exchange-splitting, and $\rho_\sigma=-(+)1$ for up (down)
spins.~\cite{rf:Kashiwaya,rf:Kawabata} The degree of the spin-filtering is characterized
by the spin-filtering efficiency $P=\left| t_\uparrow - t_\downarrow
\right|/(t_\uparrow+t_\downarrow)$, where $t_\sigma=1/\left[  1+ (Z+ \rho_\sigma S)^2
\right]$ is the transmission probability of the SF barrier for spin $\sigma$ with $m$,
$k_F$, $Z \equiv mV / k_F$, and $S \equiv m U / k_F$ being the mass of
electrons, the Fermi wave number, the normalized spin-independent and -dependent
potential barrier-height, respectively. For a perfect SF ($t_\uparrow > 0$ and
$t_\downarrow=0$), we get $P=1$.

The normal-reflection probability $B_\sigma$ and the Andreev-reflection probability $A_\sigma$ of the junction are obtained by solving the Bogoliubov-de Gennes equation
\begin{align}
\left[
\begin{array}{cc}
H_0-\rho_\sigma U \delta (x) & \Delta(x) \\
\Delta^* (x)  & -H_0+\rho_\sigma U \delta (x)
\end{array}
\right]
\Phi_\sigma (x)=
E
\Phi_\sigma (x)
,
\end{align}
together with the appropriate boundary conditions at the SF barrier ($x=0$),~\cite{rf:Kashiwaya}
where $H_0$ is the spin-independent part of the single-particle Hamiltonian, $i. e.$, $H_0= -\nabla^2/2 m+V \delta(x)-\mu_F$, $\Delta(x)=\Delta (T) e^{i \phi} \Theta(x)$ is a pair potential [$\phi$ is the phase of the pair potential and $\Theta(x)$ is the Heaviside step function], $\Phi_{\sigma} (x)$ is the eigenfunction, and the eigenenergy $E$ is measured from the chemical potential $\mu_F$.

%%%%%%%%%%%
%%%%%%%%%%%
%%%%%%%%%%%
%%%%%%%%%%%
\begin{figure}[t]
\begin{center}
\includegraphics[width=8.2cm]{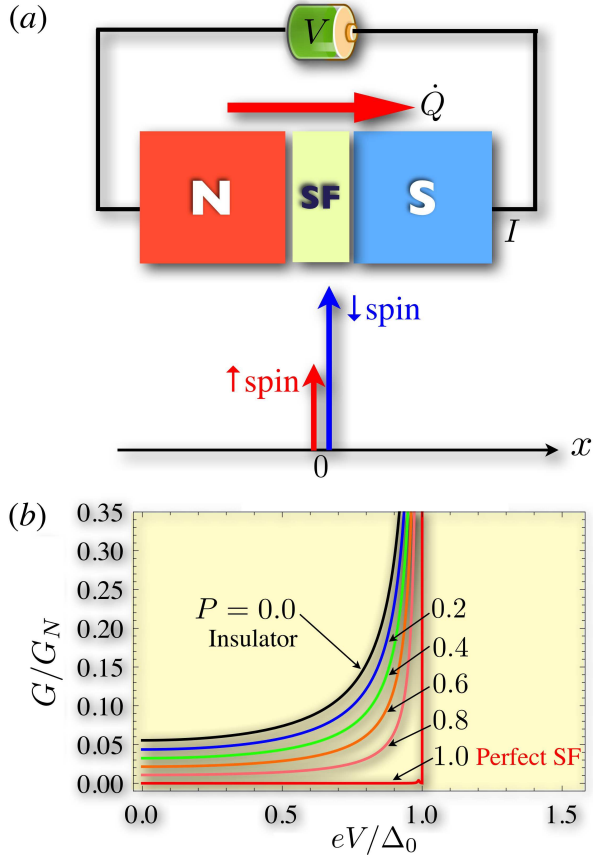}
\end{center}
\caption{(a) The schematic of a clean normal-metal/spin-filter/superconductor (N/SF/S)
cooler and the delta-function model of an SF barrier which allows the spin-selective
tunneling and the suppression of the Andreev reflection.
(b) The differential conductance
$G$  for various values of $P$ vs the bias voltage $V$ at $T=0$K for an N/SF/S
junction with $t_\uparrow=0.1$. $G_N$ is the conductance of a N/SF/N junction, $\Delta_0$
is the superconducting gap at $T=0$K, and $P$ is the spin-filtering efficiency.
 }
\label{fig1}
% \vspace{-6mm}
\end{figure}
%%%%%%%%%%%
%%%%%%%%%%%
%%%%%%%%%%%
%%%%%%%%%%%

 We first focus on  the spin-dependent charge-transport of the junction and address  the suppression of the Andreev reflection by the spin-filtering effect.
The voltage $V$ dependence of the differential conductance $G$ of the system can be calculated from the Blonder-Tinkham-Klapwijk formula,~\cite{rf:Blonder} 
$
G =(e^2/2 \pi  )
\sum_{\sigma=\uparrow,\downarrow}
\left.
\left(
  1- B_\sigma + A_\sigma
\right)
\right|_{E=eV}
$.
In Fig.~1(b) we plot the spin-filtering efficiency $P$ dependence of $G/G_N$ vs $eV/\Delta_0$ for a junction with $t_\uparrow=0.1$ at $T=0$K, where
$G_N =( e^2/2 \pi ) (t_\uparrow+t_\downarrow)$ is the conductance of a N/SF/N junction and $\Delta_0\equiv\Delta(T=0$K).
When $P$ is increased, the sub-gap conductance for $|eV| \le \Delta_0$  is largely reduced.~\cite{rf:Kashiwaya}
Importantly if $P=1$, the Andreev reflection is completely inhibited, indicating that the SF would suppress the unwanted Andreev Joule heating.

%%%%%%%%%%%
%%%%%%%%%%%
%%%%%%%%%%%
%%%%%%%%%%%
\begin{figure}[b]
\begin{center}
\includegraphics[width=8.7cm]{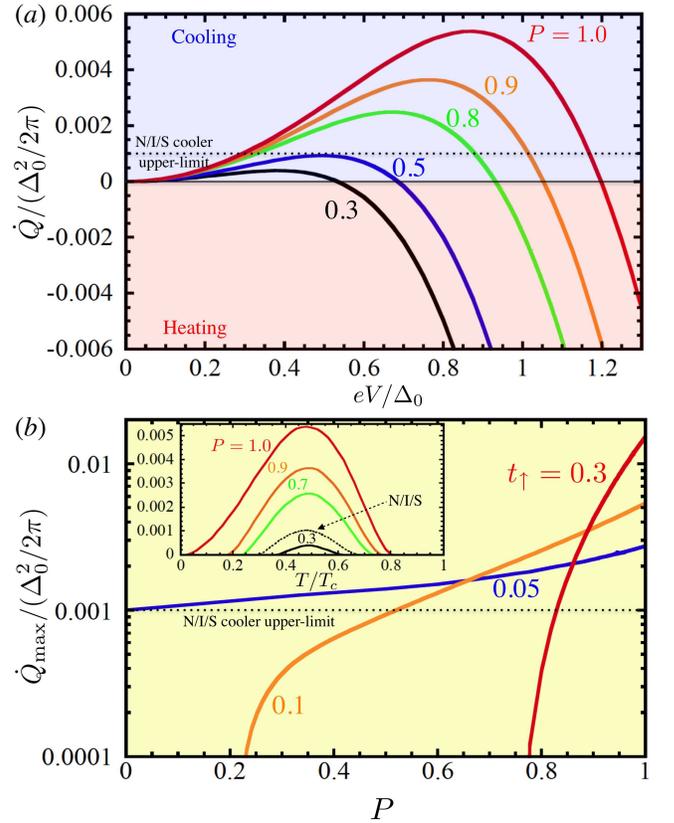}
\end{center}
\caption{(a) The cooling power ${\dot Q}$ vs the bias voltage $V$  of a clean N/SF/S refrigerator with $t_\uparrow=0.1$ at $T=0.5T_c$ for several spin filtering efficiencies $P$.
(b) The maximum cooling power $\dot{Q}_\mathrm{max}$ as a function of $P$ at $T=0.5T_c$ for several values of $t_\uparrow$.
The dotted line is the theoretical upper-limit of $\dot{Q}_\mathrm{max}  \approx 0.001 (\Delta_0^2/2 \pi )$ for N/I/S refrigerators, which is achieved in the case of $T/T_c \approx 0.5$ and $t_\uparrow=t_\downarrow \approx 0.05$.
Inset: The temperature $T$ dependence of $\dot{Q}_\mathrm{max}$ for several values of $P$.
The dashed curve shows $\dot{Q}_\mathrm{max}$ for an N/I/S refrigerator with $t_\uparrow=t_\downarrow = 0.05$.
 }
\label{fig2}
\end{figure}
%%%%%%%%%%%
%%%%%%%%%%%
%%%%%%%%%%%
%%%%%%%%%%%

To see the benefit of the spin-filtering effect on the electron cooling,  we calculate the cooling power (the amount of heat extracted from N to S) by using the Bardas and Averin formula,~\cite{rf:Bardas}
$
\dot{Q}
=
( e/ \pi )
\sum_{\sigma=\uparrow,\downarrow}
\int_{-\infty}^\infty d E
[ E ( 1- B_\sigma - A_\sigma )
-eV ( 1-B_\sigma +A_\sigma)
]
[ f(E-eV) -f(E)]
$,
where $f(E)$ is the Fermi-Dirac distribution-function.
The positive (negative) $\dot{Q}$ means the cooling (heating) of N.
In the calculation, we have determined $\Delta(T)$ by solving the BCS gap equation numerically.
The $P$ dependence of the cooling power $\dot{Q}$ vs the bias voltage $V$ for $t_\uparrow=0.1$ is shown in Fig.~2(a).
We have assumed that $T=0.5 T_c$, where $T_c$ is the superconducting transition temperature.
If we increase $P$, the cooling power $\dot{Q}$ is largely enhanced.
This  result is  attributed to the suppression of the Andreev reflection and the undesirable Andreev Joule heating.
Therefore we can conclude that the spin-filtering effect boosts dramatically  the cooling power with respect to  conventional N/I/S coolers.
In Fig.~2(b) we plot the spin-filtering efficiency $P$ dependence of the cooling-power
$\dot{Q}_\mathrm{max}$  at $T = 0.5 T_c$ and the optimal bias voltage $V=V_\mathrm{opt}$
in which $\dot{Q}$ is maximized as a function of $V$. The maximum
cooling-power $\dot{Q}_\mathrm{max}$ can be achieved in the case of the perfect SF
($P=1$) because of the complete suppression of the Andreev reflection. Notably for the
case of a large $t_\uparrow=0.3$ and $P=1$, the amount of heat extracted from N
can be about a factor of 15 larger than the theoretical upper-limit of
$\dot{Q}_\mathrm{max}$ for conventional N/I/S coolers [$\dot{Q}_\mathrm{max} \approx
0.001 (\Delta_0^2/2\pi )$], which can be achieved for $t_\uparrow =t_\downarrow
\approx 0.05$ and $T/T_c \approx 0.5$ [see the dotted line in Fig.~2(b)].~\cite{rf:Bardas}
It is crucial
to note that even for small $P$ values [$e.g.$, $P>0.0$ for $t_\uparrow =0.05$],
$\dot{Q}_\mathrm{max}$ overcomes the upper limit of N/I/S coolers. By calculating the
temperature dependence of $\dot{Q}_\mathrm{max}$, we also found that the
$\dot{Q}_\mathrm{max}$ is maximized at around $T \approx 0.5 T_c$ irrespective of the
value of $P$ [see the inset in Fig.~2(b)].

%
%
%
%\section{Diffusive junction}
%
%
%

To discuss the practical applicability of SF-based coolers, next we consider a more   {\it realistic  diffusive}  N/SF/S junction in which
 the elastic scattering length $l \ll \xi$, where $\xi=\sqrt{D/2\Delta}$ is the superconducting coherence length and $D$ is the diffusion coefficient 
 (in the following for simplicity we assume the same $D$ in the whole structure). 
 We assume that the SF is a tunnel barrier and the N reservoir is infinite along the $x$ direction [see the inset in Fig.~3(a)].

The cooling power is given by the expression 
$\dot{Q}=- I^E - I V$, where  $I$  and  $I^E$ are the charge and energy currents respectively. 
 One can express $\dot{Q}$ in terms of the
contributions from the single-particle ($|E|>\Delta$) and the Andreev ($|E|<\Delta$)
processes~\cite{rf:Vasenko} as $\dot{Q}= \dot{Q}_1 + \dot{Q}_A$, where $\dot{Q}_1 = -
I^E_1 - I_1 V$ and $\dot{Q}_A =-I^E_A - I_A V$. Here $I^E_{1(A)}$ and $I_{1(A)}$ are
respectively the single-particle  (Andreev) energy-current and the single-particle
(Andreev) charge-current. { Importantly,} the contribution of the Andreev processes
to the energy current vanishes, $I^E_A = 0$. Therefore the Andreev process contributes
only to the Joule heating ({\it i.e.} $\dot{Q}_A =- I_A V$), which is fully released in
the N electrode and leads to a severe reduction of the cooling power.

In order to compute the charge and energy   currents
through the junction we use  the quasiclassical Green's function (GF) technique. These  
 are given  by\cite{Belzig} $I = (\sigma_N \mathcal{A}/8e) \int_0^{\infty} dE \tr  \left[ \tau_3
\check{J}^K \right]$ and $I^E =(\sigma_N \mathcal{A}/8 e^2) \int_0^{\infty} dE E \tr
\left[ \tau_0 \check{J}^K \right]$, where $\sigma_N$ is the conductivity of N, $\mathcal{A}$ is the junction area, $\tau_0$ is the unit matrix,  $\tau_3$ is the Pauli matrix in the Nambu space, 
 $\breve{J} = \breve{G}\partial_x \breve{G}$ is the matrix current, and $\breve G$ is the quasiclassical GF which is a 
 $8\times 8$matrix in the Keldysh $\times$ Nambu $\times$ spin space. 
 In order to compute the currents $I$ and $I_E$ we  assume a large SF barrier resistance $R_N$, such that 
 $\xi/(R_N \sigma_N \mathcal{A})\ll 1$.
 This assumption allows for a linearization of the Usadel equation~\cite{Usadel}  in the normal metal.~\cite{rf:Ozaeta}
 The effect of the spin-filter barrier is included in the boundary conditions for the Usadel equation which has been 
  recently derived in [\onlinecite{rf:Bergeret2}].  According to the latter reference the Andreev reflection is proportional to $\sqrt{1-P}$.
   Thus,  by increasing $P$ we expect   a suppression of  the unwanted Andreev Joule
heating, {\it i.e.} an enhancement of the cooling power, as it turns our from our quantitative analysis below.

%%%%%%%%%%%
%%%%%%%%%%%
%%%%%%%%%%%
%%%%%%%%%%%
\begin{figure}[t]
\begin{center}
\includegraphics[width=7.70cm]{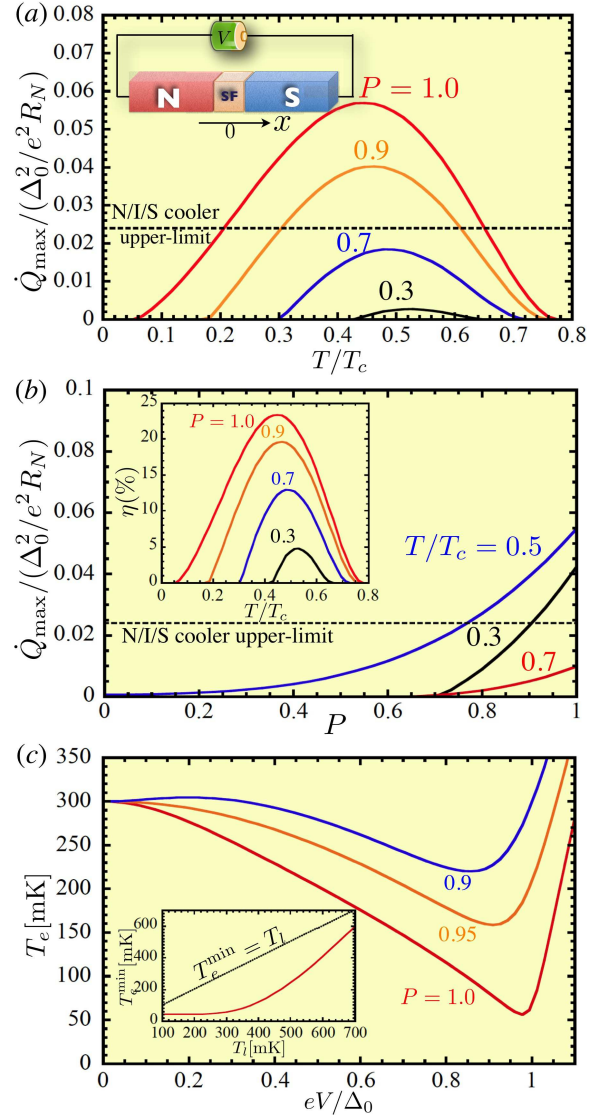}
\caption{
(a) The maximum cooling power $\dot{Q}_\mathrm{max}$  as a function of the temperature $T$ for a diffusive N/SF/S junction shown in the  inset and different values of $P$.
(b) The dependence of  $\dot{Q}_\mathrm{max}$ on  $P$ for $T/T_c$=0.7 (red), $T/T_c$=0.5 (blue), and $T/T_c$=0.3 (black).
Other parameters are  $\mathcal{A}=1 \ \mu \mathrm{m}^2$, $\sigma_N= 0.015 \ (\mu \Omega \mathrm{cm})^{-1}$ and $R_N=1.0 \ \mathrm{M} \Omega$.
Inset: The refrigeration efficiency  $\eta$  as a function of  $T$.
The horizontal dotted line in (a) and (b) is the theoretical upper-limit of $\dot{Q}_\mathrm{max}$ for diffusive N/I/S junctions, which is achieved in the case of $\mathcal{A}=1 \ \mu \mathrm{m}^2$, $\sigma_N= 0.015 \ (\mu \Omega \mathrm{cm})^{-1}$, $R_N=2.7 \ \mathrm{M} \Omega$, and $T/T_c=0.45$.
(c) The electron temperature $T_e$ as a function of $V$  for different values of $P$ and  $T_l=T_S=300$ mK.
Inset: the minimum electron temperature $T_e^\mathrm{min}$ as a function of the starting lattice temperature $T_l$ ($=T_e$ at $V=0$) for $P=1$. 
As a reference we show $T_e^\mathrm{min}=T_l$ line.
We have chosen  $\Sigma= 2 \times10^{-9} \ \mathrm{WK}^{-5} \mu \mathrm{m}^{-3}$, $\mathcal{V}=0.5 \ \mu \mathrm{m}^{3}$, and $\Delta=180 \ \mu \mathrm{eV}$.
}
\label{f3} \vspace{-4mm}
\end{center}
\end{figure}
%%%%%%%%%%%
%%%%%%%%%%%
%%%%%%%%%%%
%%%%%%%%%%%

By following the procedure described above,  we compute  numerically  the cooling power $\dot{Q}$ of our system as a function of the different parameters.
We assume a junction area $\mathcal{A}=1 \ \mu \mathrm{m}^2$, a conductivity of N $\sigma_N= 0.015 \ (\mu \Omega \mathrm{cm})^{-1}$,~\cite{rf:ONeil} and $R_N=1.0 \ \mathrm{M} \Omega$.~\cite{rf:Moodera1}
In Fig.~3(a), we plot the maximum cooling power $\dot{Q}_\mathrm{max}$ as a function of temperature $T$.
As in the ballistic junction limit ${\dot Q}_{max}$ increases dramatically by increasing $P$ reaching much larger values than for a diffusive N/I/S cooler [the dotted line in Fig.~3(a) and (b)], which is achieved in the case of $\mathcal{A}=1 \ \mu \mathrm{m}^2$, $\sigma_N= 0.015 \ (\mu \Omega \mathrm{cm})^{-1}$, $R_N=2.7 \ \mathrm{M} \Omega$, and $T/T_c=0.45$.
Moreover, the window of positive values for  ${\dot Q}_{max}$  is larger for the larger $P$.  As for the N/I/S junctions, there is a maximum value of temperature, 
$T_{max} \sim 0.75 T_c$ for which cooling is achieved [{\it c.~f.} Fig.~3(a)]. This maximum value  holds for a wide range of parameters.~\cite{rf:Vasenko,rf:Ozaeta} and is due 
to the increase of the number of thermally excited quasiparticles that contribute to the Joule heat.   
By lowering the temperature from $T=T_{max}$,   the cooling power (at optimal bias $V_\mathrm{opt}$)  first increases, reaches a maximum and finally decreases due to the Joule heat produced by the Andreev processes.
At certain temperature the cooling power tends to zero, which defines the lowest  temperature for the cooling regime.

In Fig.~3(b) we show the dependence of the  cooling power on the spin-filter efficiency $P$.
For all temperatures ${\dot Q}_{max}$ increases  monotonically by increasing  $P$. 
The efficiency of a refrigerator is characterized by the ratio between the optimum cooling-power and total input power: $\eta=\dot{Q}_\mathrm{max}/I V_\mathrm{opt}$.
 The inset of Fig.~3(b) shows the temperature $T$ dependence of  $\eta$ for several $P$ values.
 For a fully polarized  SF barrier  ($P=1$), $\eta$ reaches up to  $23\%$, which is mach larger than that for the optimum N/I/S cooler ($\eta=15 \%$) and comparable to that for a half-metallic N/FM/S cooler with $P=1$.~\cite{rf:Giazotto06}
 We can then  conclude that the spin-filtering effect gives rise to highly efficient refrigeration.

 Next we evaluate the final electron temperature of the normal metal, so we need to consider the mechanism of energy transfer. In this case the electron temperature depends on  the efficiency with which heat generated in the electron population can be transferred to the low-temperature bath. Due to the voltage biasing some power is dissipated as heat, this power is supplied initially to the electrons in the metal and is transmitted to the bath by phonons. In thin films at low temperatures, the wavelength of a thermal phonon is much less than the film thickness so the temperature of the phonons is the same as the temperature of the bath, the lattice temperature $T_l$. Then the electron temperature is determined by the rate at which electrons can transfer energy to the phonons which is given by $\dot{Q}_{e-l}=\Sigma \mathcal{V} (T_e^5-T_{l}^5)$, here $\mathcal{V}$ is the volume of N, $T_{e(l)}$ is the electron (lattice) temperature and $\Sigma$ is a material dependent parameter.~\cite{rf:Giazotto06} 
The final electron temperature $T_e$ is determined by the energy-balance equation $\dot{Q}(T_e, T_{l})+\dot{Q}_{e-l}(T_e,T_{l})=0$, where we set the temperature of the superconductor ($T_S$) equal to the bath temperature $T_l$.
In order to understand the behavior of the final electron temperature, Fig.~3(c) shows $T_e$ as a function of bias voltage for 4 different $P$ in the case of the starting lattice temperature $T_l=300$ mK.
The junction parameters were taken according to experimental values:~\cite{rf:Giazotto06,rf:Giazotto02} $\Sigma= 2 \times10^{-9} \ \mathrm{WK}^{-5} \mu \mathrm{m}^{-3}$, $\mathcal{V}=0.5 \ \mu \mathrm{m}^{3}$ and assuming that Al is the superconductor ($\Delta=180 \ \mu \mathrm{eV}$).
Fig.~3(c) shows a remarkable reduction of $T_e$, as we increase the value of voltage, the temperature tends to lower until it reaches a optimum voltage ($eV \sim \Delta_0$).
We observe that the increment of $P$ reduces drastically the minimum electron temperature, {\it i.~e.} $T_e^\mathrm{min} \sim 50$ mK for $P=1$.
In the inset of Fig.~3(c) we plot the minimum electron temperature $T_e^\mathrm{min}$ vs the starting lattice temperature $T_l$.
The straight dotted line marked $T_e^\mathrm{min}=T_l$ as a reference.
The result indicates that in a wide $T$ range, we can effectively cool down the electron temperature of N.

In summary, we have proposed an electron-refrigerator based on spin-filter  barriers. Due to the suppression of the Andreev Joule heating N/SF/S junctions can achieve values for the cooling power higher than those predicted for conventional N/I/S coolers.
Refrigeration efficiency of $15\% \sim 23\%$  can be achieved by using well known spin-filters barriers, as for example EuS/Al ($P \sim 0.86$)~\cite{rf:Moodera1}, EuSe/Al ($P \sim 1$)~\cite{rf:Moodera2} and GdN/NbN junction ($P \sim 0.75$).~\cite{rf:Senapati}
We also expect a similar effect  using  spinel ferrites ($e.g.,$ NiFe$_2$O$_4$)~\cite{rf:Luders,rf:Takahashi} or manganites ($e.g.,$ LaMnO$_{3+\delta}$ and Pr$_{0.8}$Ca$_{0.2}$Mn$_{1-y}$Co$_y$O$_3$).~\cite{rf:Satapathy,rf:Harada} as spin-filters. Our results open up a way to make  efficient solid-based refrigerators for the  cooling of  practical devices, like superconducting X-ray detectors, single-photon detectors, magnetic sensors, nano electromechanical systems, and qubits.

%
%
%
%\section{Summary}
%
%
%

We would like to thank S. Kashiwaya, M. Koyanagi, and S. Nakamura for useful discussions and comments.
This work was  supported by the Topological Quantum Phenomena (No.22103002) KAKENHI on Innovative Areas, the JSPS Institutional Program for Young Researcher Overseas Visits, the European Union Seventh Framework Programme (FP7/2007-2013) under grant agreement "INFERNOS" No. 308850, the Spanish Ministry of Economy and Competitiveness under Project FIS2011-28851-C02-02,  the CSIC and the European Social Fund under JAE-Predoc program.


\begin{thebibliography}{99}
%
%
\bibitem{rf:Nahum}
M. Nahum, T. M. Eiles, and J. M. Martinis,
Appl. Phys. Lett {\bf 65}, 3123 (1994).
%
%
\bibitem{rf:Leivon}
M. M. Leivo, J. P. Pekola, and D. V. Averin,
Appl. Phys. Lett. {\bf 68},1996 (1996).
%
%
\bibitem{rf:Giazotto06}
F. Giazotto, T. T. Heikkil\"{a}, A. Luukanen, A. M. Savin, and J. P. Pekola,
Rev. Mod. Phys. {\bf 78}, 217 (2006).
%
%
\bibitem{rf:Muhonen}
J. T. Muhonen, M. Meschke, and J. P. Pekola,
Rep. Prog. Phys. {\bf 75}, 046501 (2012).
%
%
\bibitem{rf:Clark}
A. M. Clark, N. A. Miller, A.Williams, S. T. Ruggiero, G. C. Hilton, L. R. Vale, J. A. Beall, K. D. Irwin, and J. N. Ullom,
Appl. Phys. Lett. {\bf 86}, 173508 (2005).
%
%
\bibitem{rf:ONeil}
G. C. O$'$Neil, P. J. Lowell, J. M. Underwood, and J. N. Ullom,
Phys. Rev. B {\bf 85}, 134504 (2012).
%
%
\bibitem{rf:Lowell}
P. J. Lowell, G. C. O$'$Neil, J. M. Underwood, and J. N. Ullom,
Appl. Phys. Lett. {\bf 102}, 082601 (2013).
%
%
\bibitem{rf:Bardas}
A. Bardas and D. Averin,
Phys. Rev. B {\bf 52}, 12873 (1995).
%
%
\bibitem{rf:Rajauria}
S. Rajauria, P. Gandit, T. Fournier, F. W. J. Hekking, B. Pannetier, and H. Courtois,
Phys. Rev. Lett. {\bf 100}, 207002 (2008).
%
%
\bibitem{rf:Vasenko}
A. S. Vasenko, E. V. Bezuglyi, H. Courtois, and F. W. J. Hekking,
Phys. Rev. B {\bf 81}, 094513 (2010).
%
%
\bibitem{rf:Giazotto02}
F. Giazotto, F. Taddei, R. Fazio, and F. Beltram,
Appl. Phys. Lett. {\bf 80}, 3784 (2002).
%
%
\bibitem{rf:Ozaeta}
A. Ozaeta, A. S. Vasenko, F. W. J. Hekking, and F. S. Bergeret,
Phys. Rev. B {\bf 85}, 174518 (2012).
%
%
\bibitem{rf:Wolf}
S. A. Wolf, D. D. Awschalom, R. A. Buhrman, J. M. Daughton, S. von Moln\'ar, M. L. Roukes, A. Y. Chtchelkanova, and D. M. Treger,
Science {\bf 294}, 1488 (2001).
%
%
\bibitem{rf:Moodera1}
J. S. Moodera, T. S. Santos, and T. Nagahama,
J. Phys. Cond. Mat. {\bf 19}, 165202 (2007).
%
%
\bibitem{rf:Kashiwaya}
S. Kashiwaya, Y. Tanaka, N. Yoshida, and M. R. Beasley,
Phys. Rev. B {\bf 60}, 3572 (1999).
%
%
\bibitem{rf:Bergeret2}
F. S. Bergeret, A. Verso, and A. F. Volkov,
Phys. Rev. B {\bf 86}, 214516 (2012).
%
%
\bibitem{rf:Kawabata}
S. Kawabata, S. Kashiwaya, Y. Asano, Y. Tanaka, and A. A. Golubov,
Phys. Rev. B {\bf 74}, 180502(R) (2006).
%
%
\bibitem{rf:Blonder}
G. E. Blonder, M. Tinkham, and T. M. Klapwijk,
Phys. Rev. B {\bf 25}, 4515 (1982).
%
%
\bibitem{Belzig}
W.~Belzig, F.~K.~Wilhelm, C.~Bruder, G.~Sch\"{o}n, and A.~D.~Za\-i\-kin,
Superlatt. Microstruct. {\bf 25}, 1251 (1999).
%
%
\bibitem{Usadel}
K.~D.~Usadel,
Phys.~Rev.~Lett. {\bf 25}, 507 (1970).
%
%
\bibitem{rf:Moodera2}
J. S. Moodera, R. Meservey, and X. Hao,
Phys.~Rev.~Lett. {\bf 70}, 853 (1993).
%
%
\bibitem{rf:Senapati}
K.~Senapati, M.~G. Blamire, and Z.~H.~Barber,
Nat. Mater. {\bf 10}, 849 (2011).
%
%
\bibitem{rf:Luders}
U. L\"uders, M. Bibes, K. Bouzehouane, E. Jacquet, J. -P. Contour, S. Fusil, J. -F. Bobo, J. Fontcuberta, A. Barth\'el\'emy, and A. Fert,
Appl. Phys. Lett. {\bf 88}, 082505 (2006).
%
%
\bibitem{rf:Takahashi}
Y. K. Takahashi, S. Kasai, T. Furubayashi, S. Mitani, K. Inomata, and K. Hono,
Appl. Phys. Lett. {\bf 96}, 072512 (2010).
%
%
\bibitem{rf:Satapathy}
D. K. Satapathy, M. A. Uribe-Laverde, I. Marozau, V. K. Malik, S. Das, Th. Wagner, C. Marcelot, J. Stahn, S. Br\"uck, A. R\"uhm, S. Macke, T. Tietze, E. Goering, A. Fran\'o, J.-H. Kim, M. Wu, E. Benckiser, B. Keimer, A. Devishvili, B. P. Toperverg, M. Merz, P. Nagel, S. Schuppler, and C. Bernhard,
Phys. Rev. Lett. {\bf 108}, 197201 (2012).
%
%
\bibitem{rf:Harada}
T. Harada, I. Ohkubo, M. Lippmaa, Y. Sakurai, Y. Matsumoto, S. Muto, H. Koinuma, and M. Oshima,
Phys. Rev. Lett. {\bf 109}, 076602 (2012).
%
%
\end{thebibliography}
\end{document}